\documentclass[prb,preprint,showpacs,superscriptaddress]{revtex4}
\usepackage{graphicx}

\begin{document}
\title{ZnSe/GaAs(001) heterostructures with defected interfaces:\\
 structural, thermodynamic  and electronic properties}

\author{A. Stroppa}
\email{astroppa@ts.infn.it}
\affiliation{Dipartimento di Fisica Teorica, Universit\`a di
Trieste,\\ Strada Costiera 11, I-34014 Trieste, Italy}
\affiliation{INFM DEMOCRITOS National Simulation Center, Trieste, Italy}
\author{M. Peressi}\texttt{\texttt{}}
\email{peressi@ts.infn.it}
\affiliation{Dipartimento di Fisica Teorica, Universit\`a di
Trieste,\\ Strada Costiera 11, I-34014 Trieste, Italy}
\affiliation{INFM DEMOCRITOS National Simulation Center, Trieste, Italy}
\date{\today}

\begin{abstract}
We have performed accurate \emph{ab--initio} pseudopotential
calculations for the structural and electronic properties of
ZnSe/GaAs(001) heterostructures with interface
configurations accounting for charge neutrality prescriptions.
Beside the simplest configurations with atomic interdiffusion
 we consider also some configurations characterized by
As depletion and cation vacancies, motivated by the
recent successfull growth of ZnSe/GaAs
pseudomorphic structures with minimum stacking fault density
characterized by the presence of a defected (Zn,Ga)Se alloy
in the interface region.
We find  that---under particular thermodynamic
conditions---some defected configurations
are favoured with respect to undefected ones
with simple anion or cation mixing, and that
 the  calculated band offsets for
some defected structures are compatible with those measured.
Although it is not possible to extract indications about the
precise interface composition and vacancy concentration, our
results  support the experimental indication of (Zn,Ga)Se defected
compounds in high-quality ZnSe/GaAs(001) heterojunctions with low
native stacking fault density. The range of measured band offset
suggests that different atoms at interfaces rearrange, with
possible presence of vacancies, in such a way that not only local
charges but also ionic dipoles are vanishing.

\end{abstract}

\pacs{PACS: 73.40.Kp, 73.20.-r, 68.35.-p}

\maketitle

%%%%%%%%%%%%%%%%%%%%%%%%%%%%%%%%%%%%%%%%%%%%%%%%%%%%%%%%%%%%%%%%%%%%%%%%%%%%%%%%
\section{Introduction}
II-VI/III-V heterostructures, whose prototype is ZnSe/GaAs,
fabricated by molecular beam epitaxy (MBE) are
important for electronic devices such as
blue-green emitters~\cite{bluegreen} in optoelectronics,
spin-transistors~\cite{spintrans1,spintrans2} and spin
filters~\cite{spinfilter} in spintronics.
For all the proposed applications, it is crucial to have a control of
the structural quality of these interfaces and, in particular, of the
 native stacking fault (SF) defect density which causes serious
 device degradation.

The SFs density
in II-VI/III-V interface is determined by the growth conditions.
It has been suggested that the reduction of stacking fault density in the
ZnSe/GaAs heterostructure can be achieved by inserting a thin
low-temperature-grown ZnSe buffer layer between the high-temperature-grown
ZnSe epilayer  and the GaAs substrate, thus obtaining SF density as low as
$\sim$5.4$\times$10$^{4}$ cm$^{-2}$ ~\cite{LTZnSe}. More recently, two
procedures have been found to work in reducing the SF density even below
 10$^{4}$ cm$^{-2}$ (thus providing a very high quality system),
yielding quantitatively similar defected densities
 and  qualitatively similar interface compositions
and band alignments~\cite{Franciosi}.

A part from this experimental evidence, the microscopic mechanisms
that control the native defect density in II-VI/III-V structures
remain still controversial. What is noteworthy is that from an
accurate characterization of the samples with minimum stacking
fault density~\cite{Franciosi} there is experimental evidence of
the formation of a ternary (Zn,Ga)Se alloy of variable composition
with a substantial concentration of cation
vacancies.~\cite{ColliJAP2004} Because of vacancies, its average
lattice parameter is smaller than the one of GaAs and ZnSe, and
therefore this alloy is under tensile biaxial strain when
epitaxially grown on GaAs substrates, accumulating a non
negligible elastic energy. Evidence of formation of ordered binary
defected compounds such as Ga$_{2}$Se$_{3}$ or interface layers
with vacancies arranged with other symmetry at ZnSe/GaAs
interfaces has also been
reported.~\cite{LiAPL1990,reviewcompounds} The defected compound
Ga$_{2}$Se$_{3}$ has also intentionally been grown epitaxially on
GaAs~\cite{LiJVSTB1997} and, very recently, on Si(001)
substrates~\cite{OhtaPRL2005} for its utilization in
optoelectronics.

The driving  mechanism
for the formation of this defected alloy at ZnSe/GaAs interfaces is still
unknown and, moreover, it is not clear whether defected interfaces could
be favourite with respect to the simplest cases of
cation and/or anion mixed interfaces without vacancies.

To this aim we have performed a comparative study based on \emph
{ab-initio} local-density-functional pseudopotential approach of
ZnSe/GaAs (001) heterostructures, including a few selected
interface configurations with cation vacancies. Since charged
interfaces are energetically unstable with respect to the
interdiffusion of atoms across the
interface~\cite{reconstr1,reconstr2,unstable3,unstable4,unstable5,gaasznse}
we consider  interfaces satisfying the charge neutrality
condition. Our comparative study will address their structural and
electronic properties as well as their formation energy.\\
 This work is organized as follows: in the next section we describe
the theoretical and computational approach; in Sect. III we describe
the selected interface morphologies; in Sect. IV and V  we
 report our results for the structural properties and relative stability; Sect. VI is devoted to
 the discussion of the electronic properties in terms of  Density of States (Sect. VI-A) and
  band alignments (Sect. VI-B); finally, in Sect. VII we draw our conclusions.
%%%%%%%%%%%%%%%%%%%%%%%%%%%%%%%%%%%%%%%%%%%%%%%%%%%%%%%%%%%%%%%%%%%%%%%%%%%%%%%%

\section{Theoretical and computational method}
Our calculations are performed within the density functional theory
framework using
the local density approximation for the exchange-correlation
functional~\cite{lda1,lda2} with  state-of-the-art first-principles
pseudopotential
self-consistent calculations.~\cite{PWSCF}
 The wave functions are expanded onto
a plane-wave basis set with a kinetic energy cutoff of
 20 Ry: we carefully check that all the
relevant bulk properties of the binary ZnSe and GaAs
 compounds are well converged.
 The Zn-3d electrons are taken into account only via the nonlinear core
 correction which has been shown to  give reliable results for ZnSe,
 as discussed in
 Ref. ~\onlinecite{gaasznse}.
The integration over the Brillouin zone is performed using
the special {\bf k}-point technique with a  (6,6,6)
 Monkhorst-Pack mesh for the FCC cell and corresponding meshes for the
various supercells.

The theoretical  lattice constant of GaAs (ZnSe) is 1.8 \%  (1.4 \%) smaller
than the experimental lattice constant a$^{GaAs}_{exp}$=5.65 {\AA}~\cite{Landolt}
(a$_{exp}^{ZnSe}$=5.67 {\AA}~\cite{Landolt})  but the small
relative lattice mismatch is well reproduced ($\leq 0.8$ \%). In
our calculations, we neglect this tiny lattice mismatch and we
fixed the in-plane lattice constant to the theoretical one of GaAs
which is typically the substrate in experimental samples.
The ionic degrees of freedom are fully taken into account by
optimizing  the atomic positions via a total-energy and
atomic-force minimization~\cite{forces} with a threeshold of
 1 mRy a.u.$^{-1}$ for atomic forces and of 0.01 mRy
for total energy changes between two consecutive
relaxation steps.

The interfaces are modeled by tetragonal supercells with periodic
boundary conditions and contain a slab for each constituent
material.
We study the
relative stability by considering the
interface formation energy:~\cite{Eform}
\begin{equation}\label{eq:eform}
2E^{intf}_{f}=\frac{(E_{supercell}-\sum_{i=1}^{N_{species}}
n^{i}\mu^{i})}{m n}
\end{equation}
 where $E_{supercell}$ is the calculated total energy of the
supercell, $N_{species}$ is
  the number of the chemical species involved
(which are 4 in the present case), $n^{i}$ the
 number of atoms of the species $i$ and $\mu^{i}$ is the corresponding
 chemical potential;  $(m \times n)$ is the
reconstruction (the in-plane periodicity of the interface), so that
 $N=mn$ is the number of atoms in the two-dimensional
interface unit cell and  $E^{intf}_{f}$ refers to a (1$\times$1)
 interface area.
In general $E^{intf}_{f}$ refers to a
 \emph{mean}
 value of the formation  energy of the two possibly inequivalent interfaces
 present in each supercell.
We calculated the bulk chemical potentials considering the elemental
 forms  of Zn (hcp),~\cite{bulkform} Se (trigonal),~\cite{bulkform}
 Ga (orthorhombic),\cite{bulkform} As (trigonal),~\cite{bulkform}
  GaAs and ZnSe (cubic).~\cite{Landolt}
We employed the Gaussian smearing technique~\cite{Smearing} for
Brillouin integration of metallic systems. For each bulk elemental
form we have chosen the proper $k$-point mesh which gives well
converged values of the corresponding chemical potential, with a
numerical uncertainty less than  $\approx$20 meV per bulk formula
units.
 The calculated heat of
formation of bulk ZnSe and GaAs are $-$1.61 eV and $-$0.80 eV
respectively which well compare with the experimental values
($-$1.67 eV and $-$0.84 eV):\cite{Landolt,expheat} this
 indirectly tests the reliability of our
calculated elemental chemical potentials. They also agree with
previous theoretical
calculations.\cite{gaasznse,AZalloy2003,ref1,ref2,ref3}

%%%%%%%%%%%%%%%%%%%%%%%%%%%%%%%%%%%%%%%%%%%%%%%%%%%%%%%%%%%%%%%%%%%%%%%%%%%%%%%
\section{Interface structures}\label{interfaces}

We discuss in this Section the selected  interface morphologies
 that we use as simplest models to describe
 the defected ZnSe/GaAs (001) junctions. Based on experimental suggestions
about the formation of
(Zn,Ga)Se compound in a As-depleted region\cite{ColliJAP2004}
and for the sake of simplicity
 we consider only cation vacancies and
 we  exclude the possibility of other defects such as
interstitials or antisites.

 Simple electrostatic considerations, already pointed out by
 Harrison along time ago,\cite{reconstr1} suggest that no ionic charge
accumulation can occur at the junction since uncompensated charges
would set up an electric field extending throughout the overlayer,
with an energy accumulation clearly unfavourable as long as the overlayer
thickness increases.\cite{note1}
The charge neutrality condition at the interface could be expressed in terms
of ionic charges by calculating their {\em macroscopic average} density profile along
the direction of the junction and imposing that possible deviations
are compensated.\cite{review} Alternatively, it  can be formulated also in terms
of {\em bond} charge compensation.\cite{reconstr2}
 In general, at interfaces between
heterovalent constituents there are ``wrong'' chemical bonds with
either more (``donor  bond'') or less (``acceptor bond'') than two
electrons.\cite{wrong} A practical rule for the \emph{compensation}
of  donor and acceptor bonds can be derived as follows. In the
zincblende structure between two consecutive atomic planes $i$ and
$i+1$ in the [001] direction there are two bonds for an area of
$a^2_0/2$, where $a_0$ is the bulk lattice constant, with two
electrons each one and hence with a total of four electrons. These
electrons come from the contributions $\rm Z_i/2$ and $\rm
Z_{i+1}/2$, where $\rm Z_i$ is the ionic charge of the atoms in the
$i$ plane. $\rm Z_{i}$ and $\rm Z_{i+1}$ are 3 and 5(2 and 6) in
GaAs(ZnSe) and therefore trivially
\begin{equation}
\langle \rm Z_{i,i+i}\rangle=(Z_{i}+Z_{i+1})/2=4,\label{ruleofthumb}
\end{equation}
  where
$\langle \rm Z_{i,i+1}\rangle$ is the  average of bond electrons
between the two corresponding consecutive atomic planes. The
\emph{compensation} of donor and acceptor bonds requires that
Eq.~\ref{ruleofthumb} holds also in the interplanar region at the
interface, with
 $\rm Z_i$ referring in general to the {\em average} ionic charge per atom in the
$i$ plane.
If not satisfied over a pair of consecutive atomic planes, the condition
must be fullfilled by compensation with neighboring bonds, i.e. in
neighboring pairs of atomic planes.

%%%

We first consider, for the sake of example and comparison, the simplest
interface morphologies with  atomic intermixing limited to one or two atomic
planes, without vacancies and satisfying the charge neutrality,
already addressed in Refs.~\onlinecite{unstable4,gaasznse}. The
composition profiles for these simple cases are:
\begin{itemize}%\small
\item[{\em 1C}:] 1-plane %50\%-50\%
Cation-mixed:
  $\cdots$ -Ga-As-(Ga$_{\frac{1}{2}}$\ Zn$_{\frac{1}{2}}$)-Se-Zn-$\cdots$
\item[{\em 1A}:] 1-plane
%50\%-50\%
Anion-mixed: $\cdots$
-As-Ga-(As$_{\frac{1}{2}}$\ Se$_{\frac{1}{2}}$)-Zn-Se-$\cdots$
\item[{\em 2CA-GaSe}:] 2-planes Cation-Anion mixed, Ga and Se
rich:\\
%\\ Zn-Ga 25\%-75\% and Se-As 75\%-25\%:
$\cdots$ -Ga-As-(Zn$_{\frac{1}{4}}$\ Ga$_{\frac{3}{4}}$)-(Se$_{\frac{3}{4}}$\
As$_{\frac{1}{4}}$)-Zn-Se-$\cdots$
\item[{\em 2CA-ZnAs}:] 2-planes Cation-Anion mixed, Zn and As
rich:\\
%\\ Se-As 25\%-75\%  and Zn-Ga 75\%-25\%:
$\cdots$-As-Ga-(Se$_{\frac{1}{4}}$\
As$_{\frac{3}{4}}$)-(Zn$_{\frac{3}{4}}$\
Ga$_{\frac{1}{4}}$)-Se-Zn-$\cdots$
\end{itemize}
The profiles of the average ionic charges $\rm Z_i$ on the atomic
planes are:
\begin{itemize}
\item[{\em 1C}:] $\cdots - 3 - 5 - 2.5 - 6 - 2 - \cdots$
\item[{\em 1A}:] $\cdots - 5 - 3 - 5.5 - 2 - 6 - \cdots$
\item[{\em 2CA-GaSe}:] $\cdots - 3 - 5 - 2.75 - 5.75 - 2 - 6 - \cdots$
\item[{\em 2CA-ZnAs}:] $\cdots - 5 - 3 - 5.25 - 2.25 - 6 - 2 - \cdots$
\end{itemize}
and the sequences of  the average bond electrons in the interplanar
spaces $\langle \rm Z_{i.i+1} \rangle$ are:
\begin{itemize}
\item[{\em 1C}:] $\cdots - 4 - 3.75 - 4.25 - 4 - \cdots$
\item[{\em 1A}:] $\cdots - 4 - 4.25 - 3.75 - 4 - \cdots$
\item[{\em 2CA-GaSe}:] $\cdots - 4 - 3.875 - 4.25 - 3.875 - 4 - \cdots$
\item[{\em 2CA-ZnAs}:] $\cdots - 4 - 4.125 - 3.75 - 4.125 - 4 - \cdots$
\end{itemize}

 It is easy to check that for all these interfaces Eq.~\ref{ruleofthumb}
is satisfied by a compensation occurring  at most over three
interplanar spacings.
 The latter interfaces  have no ionic dipole,
as it would be the case for the abrupt, nonpolar (110) interface.
The  interfaces 1C, 1A and 2CA-ZnAs, 2CA-GaSe are related by a
simultaneous exchange of Ga with As and Zn with Se.

Introducing cation vacancies, many other different morphologies
compatible with the charge neutrality condition are possible.
We generalize the case of 1-plane Cation-mixed interface
by considering the possibility of another Cation-mixed plane
with vacancies as follows:
 $$\cdots -Ga-As-(Ga_{x}Zn_{y}V_{1-x-y})-Se-(Ga_{w}Zn_{z}V_{1-w-z})-Se-Zn-
 \cdots$$
 where V indicate vacancies and $x$, $y$, $w$, $z$ are the
in-plane atomic concentrations of the
 corresponding atomic species, with the obvious condition:
\begin{equation}
 x+y\leq1 \ \textrm{and} ~ w+z\leq1. \label{charge-neutr2}
\end{equation}

  For this case, the sequence of the average ionic charges
 $\rm Z_i$ on the atomic planes is:
 $$\cdots -3-5-(3x+2y)-6-(3w+2z)-6-2-
 \cdots$$
and the corresponding sequence of $\langle \rm Z_{i,i+1}\rangle$ as
defined above is:
$$\cdots -4-\frac{5+3x+2y}{2}-\frac{3x+2y+6}{2}-\frac{6+3w+2z}{2}
-\frac{3w+2z+6}{2}-4-
 \cdots$$

 The charge neutrality
condition, that is Eq.~\ref{ruleofthumb}, applied to this
composition profile gives:
\begin{equation}
6(x+w)+4(y+z)=9 \label{charge-neutr1}
\end{equation}

We note that Eq.~\ref{charge-neutr1} is symmetric under  the
exchange of $x\leftrightarrow w$ and/or $y\leftrightarrow z$. In
the following we focus only on the pairs of  \emph{complementary}
solutions related by \emph{both} $x\leftrightarrow w$ and
 $y\leftrightarrow z$
exchange and satisfy Eq.~\ref{charge-neutr2}:  this corresponds to
a \emph{swap} of the cation planes across
the Se plane.\\

First,  we are going to consider systems with a high (H) local
concentration of vacancies: in particular
 an interface plane with 50\% concentration of vacancies,
a configuration not very realistic but that can be easily
simulated with  small supercells.
This case corresponds to $x=\frac{1}{2}$, $w=1$, $y=z=0$
 and the complementary solution, with the following composition
profiles:
\begin{itemize}
\item[{\em 1HV}:]   $\cdots-Ga-As-Ga_{\frac{1}{2}}-Se-Ga-Se-Zn-\cdots$
\item[{\em 1HV-swap}:]  $\cdots-Ga-As-Ga-Se-Ga_{\frac{1}{2}}-Se-Zn-\cdots$
\end{itemize}

These profiles are schematically shown in Fig. 1
together with a sketch of the bare ionic charge profile, averaged
over two consecutive atomic planes and normalized to a bulk
zincblende cell, indicating the extension and the value of the
interface dipole.

Next, we consider the solutions  with $z$=1 and $w$=0
which reduce to
the case of  vacancies just confined in a single plane between GaAs and ZnSe.
 Out of the possible cases, we select  the one
with $x=\frac{5}{6}$ and $y=0$. Considering also the swap across the Se plane,
we have the following composition profiles:
\begin{itemize}
\item[{\em 1V}:] $\cdots-Ga-As-Ga_{\frac{5}{6}}-Se-Zn-Se-Zn-\cdots$
\item[{\em 1V-swap}:] $\cdots-Ga-As-Zn-Se-Ga_{\frac{5}{6}}-Se-Zn-\cdots$
\end{itemize}
In terms of ionic charges, each Zn is equivalent to ${\frac{2}{3}}$
of Ga, so that if we substitute Zn with
  Ga$_{\frac{2}{3}}$ in the composition profiles of the
possible neutral interfaces we still maintain the charge neutrality condition
and we have:
\begin{itemize}
\item[{\em 2V}:]
$\cdots-Ga-As-Ga_{\frac{5}{6}}-Se-Ga_{\frac{2}{3}}-Se-Zn-\cdots$
obtained from 1V;
\item[{\em 2V-swap}:]
$\cdots-Ga-As-Ga_{\frac{2}{3}}-Se-Ga_{\frac{5}{6}}-Se-Zn-\cdots$
obtained from 1V-swap.

\end{itemize} These interfaces have two
planes with vacancies and correspond to $x={\frac{5}{6}}$,
$w={\frac{2}{3}}$, $y=z=0$ and the complementary case
$x={\frac{2}{3}}$, $w={\frac{5}{6}}$, $y=z=0$.

All the interfaces considered above are characterized by the
presence of Ga-Se and/or Zn-As bonds, which  are not present in the
bulk constituents. Considering that recent experimental results do
not indicate the presence of Zn--As bonds~\cite{ColliJAP2004} we
focus here mainly on structures with Ga--Se bonds and we include for
comparison only one with Zn--As bonds, namely the one labelled as
``1V-swap''. Remarkably, in the entire range of the chemical
potentials involved, the interface having  Zn-As bonds is
thermodynamically unstable with respect to the complementary one
(1V) which has the same stoichiometry but no Zn-As bonds (see
Section\ref{thermo}).

\section{Structural properties}

Because of the cation vacancies,
lattice distortions are sizeable  and affect mainly
 the anion sublattice.
Below, we briefly discuss the structural properties of the different
configurations studied. Lattice distortions have non negligible effects on
electronic properties and stability of the system, which will be discussed
in the next Section.

\subsection{Interfaces  with 50\% vacancies layer}
Both 1HV and 1HV-swap supercells are characterized by a huge
local concentration of cation vacancies, having
 a  plane with 50\% of vacancies. The supercells considered have 2 atoms per
 layer and  20 atomic layers (the total length
 of the supercell is 33.27 \AA).
 Atomic  relaxations are important in the interface region, and
result in sizeable variations of the bond lengths---involving in
particular of acceptor and donor bonds---and interplanar distances
along the growth direction, which are reported in Fig. 2.

\subsection{Interfaces with different vacancy concentration}
The supercell describing the 1(2)V and 1(2)V-swap interfaces  have
6 atoms per layer and  20 atomic layers, with a total length of
the supercells of 27.72 \AA.
 In Fig.~\ref{fig2.ps}
 we show the average interplanar distances along the [001]
 direction. In this cases, we have a non negligible  {buckling}
of the atomic planes, in particular in the interface region, which reduces
progressively far from the interface, although propagating
also several atomic layers far away,
 much more in 2V and 2V-swap than in 1V and 1V-swap.
In Fig. 2 we
report the average interplanar distances with an error bar determined by
   the maximum
  and minimum interlayer distances between adjacent atomic layers.

Although the composition profile along the growth direction is
symmetric with respect to a middle plane in the constituent slabs,
the two interfaces in the simulation cell are not equivalent in
their 3D structure, and this justifies the lack of symmetry in the
pattern of the interatomic distances. Incidentally, we note that
the larger is the buckling in the interface region, the stronger
is the effect propagating in the bulk slabs.

\section{Thermodynamic stability of interfaces}\label{thermo}
We study the relative stability of the
interfaces using Eq.~(\ref{eq:eform}). Although the
precise values of the chemical potentials are unknown,
strongly depending on the growth process and on the local environment,
their range of variation and mutual relationships can be established in
condition of thermodynamic equilibrium. In particular:
\begin{equation}\label{eq1}
\mu_{GaAs}^{bulk}=\mu_{Ga}+\mu_{As} \textrm{ and }
\mu_{ZnSe}^{bulk}= \mu_{Zn}+\mu_{Se}
\end{equation}
 for the
equilibrium between the interface and the bulks;

\begin{equation}\label{eq2}
\mu_{i}\leq\mu_{i}^{bulk}
\end{equation}
 to exclude the formation of
precipitates (since when $\mu_{i}=\mu_{i}^{bulk}$ the gas phase
condensates to form the elemental bulk phase).  We could
 also exclude the formation of other precipitates, like
Zn$_{3}$As$_{2}$ and Ga$_{2}$Se$_{3}$, but this would only
further restrict the possible range of variations of the chemical
potentials involved, without changing the main
conclusions of our study.

By defining the heat of formation of AB compound as: $\Delta
H_{AB}=\mu_{AB}^{bulk}-\mu_{A}^{bulk}
 -\mu_{B}^{bulk}$,\cite{gaasznse} so that a negative $\Delta H_{AB}$
 means exotermic reaction, one gets the following relations:
\begin{equation}\label{boundsas}
\Delta H_{GaAs}+\mu_{As}^{bulk}\leq\mu_{As}\leq\mu_{As}^{bulk}
\end{equation}
and
\begin{equation}\label{boundsse}
\Delta H_{ZnSe}+\mu_{Se}^{bulk}\leq\mu_{Se}\leq\mu_{Se}^{bulk}.
\end{equation}
We choose $\mu_{Zn}-\mu_{Ga}$ and  $\mu_{Se}$ as basic variables,
and with some algebra we obtain:
\begin{equation}\label{boundsznga}
\mu_{Zn}^{bulk}-\mu_{Ga}^{bulk}+\Delta H_{ZnSe} \leq
\mu_{Zn}-\mu_{Ga} \leq \mu_{Zn}^{bulk}-\mu_{Ga}^{bulk}-\Delta
H_{GaAs}
\end{equation}

We can introduce new variables
\begin{equation}
\widetilde{\mu}_{Zn}-\widetilde{\mu}_{Ga}=
\mu_{Zn}-\mu_{Ga}-(\mu_{Zn}^{bulk}-\mu_{Ga}^{bulk})
\end{equation}
and
\begin{equation}\label{muse}
\widetilde{\mu}_{Se}=\mu_{Se}-\mu_{Se}^{bulk}
\end{equation}
and discuss the relative stability of the interfaces
in the ranges of variations of these variables.

We can specify Eq.~(\ref{eq:eform}) as follows:
\begin{equation}\label{eq:eform_general}
2E_{f}^{intf}=\frac{E_{supercell}-N_{Se}\mu_{ZnSe}^{bulk}
-N_{As}\mu_{GaAs}^{bulk}+(N_{Se}-N_{Zn})\mu_{Zn}+
(N_{As}-N_{Ga})\mu_{Ga}}{mn}
\end{equation}
The supercells  with no vacancies that we have considered here are
stoichiometric (N$_{Se}$=N$_{Zn}$=N$_{ZnSe}$, and
N$_{Ga}$=N$_{As}$=N$_{GaAs}$), therefore their formation energy is
independent on the chemical potentials and
Eq.~(\ref{eq:eform_general}) reduces to a constant value. The
resulting interface formation energies for 1C and 1A are nearly
degenerate, within 10 meV per ($1\times$1) interface unit cell.
The same holds for 2CA-ZnAs and 2CA-GaSe. The relaxations lower
the formation energies by 60 meV (50 meV) per interface unit cell
for 1C and 1A, whereas the effect is stronger for 2CA-ZnAs and
2CA-GaSe where the formation energy is lowered by  $\sim$ 120 meV per
(1$\times$1) interface unit cell.\\
For the other structures, from Eq.~(\ref{eq:eform_general}) we obtain:

\begin{equation}\label{eq:eformb_b1}
E_{f,1HV}^{intf}=C_{1HV}+
(\widetilde{\mu}_{Zn}-\widetilde{\mu}_{Ga})-\frac{1}{2}\widetilde{\mu}_{Se}
\end{equation}
\begin{equation}\label{eq:eformalpha2_alpha3}
E_{f,1V}^{intf}=C_{1V}+\frac{1}{3}
 (\widetilde{\mu}_{Zn}-\widetilde{\mu}_{Ga})
-\frac{1}{6} \widetilde{\mu}_{Se}
\end{equation}
\begin{equation}\label{eq:eformalpha_alpha1}
E_{f,2V}^{intf}=C_{2V}+(\widetilde{\mu}_{Zn}-\widetilde{\mu}_{Ga})-
\frac{1}{2}\widetilde{\mu}_{Se}
\end{equation}
We observe that structures related by a swap of atomic planes have the same
 $N_{i}$'s, so they have the same expressions for
 the interface formation energy: therefore Eqs.
~(\ref{eq:eformb_b1},\ref{eq:eformalpha2_alpha3},
 \ref{eq:eformalpha_alpha1}) also hold for
 the complementary structures.
The constants $C_i$  are characteristic
of the specific interface and do not depend on the length of the
supercell used: test calculations show that they change only
by $\sim$ 10 meV per (1$\times$1) interface unit cell when increasing
the length of the supercells: this is a further indication that the size of
our supercells is large enough.

As it can be seen from Eqs.~(\ref{eq:eformb_b1},
\ref{eq:eformalpha2_alpha3},\ref{eq:eformalpha_alpha1}), the
interface energy E$_{f}^{intf}$ is a linear function of
$\widetilde{\mu}_{Zn} -\widetilde{\mu}_{Ga}$ and
$\widetilde{\mu}_{Se}$ with coefficients determined by the
particular stoichiometry. It is convenient to discuss the
dependence of E$_{f}^{intf}$ on an individual variable separately, setting
 the other one equal  to its limiting value  given by
Eqs.~(\ref{boundsse},\ref{boundsznga}). In Fig.~\ref{fig3.ps}, we
show the interface formation energies as a function of
$\widetilde{\mu}_{Zn}-\widetilde{\mu}_{Ga}$ for lower (upper)
limit of $\widetilde{\mu}_{Se}$ (top) and as a function of
$\widetilde{\mu}_{Se}$ for lower (upper) limit of
$\widetilde{\mu}_{Zn}-\widetilde{\mu}_{Ga}$ (bottom). The
horizontal line corresponds to interfaces with no vacancies,
namely 1C, 1A, 2CA-ZnAs, 2CA-GaSe which appear degenerate on the
scale used. Right (left) hatched areas represent the range of
variation of formation energy for 1HV, 1HV-swap, 2V, 2V-swap (1V,
1V-swap) configurations. The lowest formation energy for right
(left) hatched area
corresponds to 1HV (1V) configuration.

Now we discuss the  limiting cases corresponding to
$\widetilde{\mu}_{Se}=\Delta H_{ZnSe}$ and
$\widetilde{\mu}_{Se}=0$, i.e. the lowest and  uppermost values
of $\widetilde{\mu}_{Se}$
 (see Fig.~\ref{fig3.ps}, top part).
The heat of formation for ZnSe determines the lower limit of
$\widetilde{\mu}_{Se}$ (see Eqs.~\ref{boundsse},\ref{muse}).
E$_{f}^{intf}$ is a function of the other chemical potential. It
can be seen that the most stable structures remain the same (1HV
and the mixed ones with no vacancies) independently on the value
of $\widetilde{\mu}_{Se}$, although the value of
$(\widetilde{\mu}_{Zn}-\widetilde{\mu}_{Ga})$ at which they have
competing energies changes a little bit. As a general trend, high
values of $\widetilde{\mu}_{Se}$ should  stabilize the formation
of defected interfaces (in Fig.~\ref{fig3.ps}, top part, compare
the Max$\{\widetilde{\mu}_{Se}\}$  respect to
Min$\{\widetilde{\mu}_{Se}\}$ case). This observation agrees
   with previous calculation dealing with GaAs(001) surfaces where
   it has been shown that for an increasing Se chemical potential
   (high limit of $\widetilde{\mu}_{Se}$) the formation of a Ga
   vacancy beneath the surface becomes energetically favourable,
   driving the surface stoichiometry towards Ga$_{2}$Se$_{3}$.~\cite{Se-reacted} On the other hand,
substantial changes occur considering the extreme cases for
$\widetilde{\mu}_{Zn}-\widetilde{\mu}_{Ga}$ (see
Fig.~\ref{fig3.ps}, bottom part).
 This indicates that the variation of $\widetilde{\mu}_{Se}$ has much less
 effect than a variation of
 $\widetilde{\mu}_{Zn}-\widetilde{\mu}_{Ga}$. This can be also
 argued from the different relative weight of the basic variables
 in
 Eqs.~\ref{eq:eformb_b1},~\ref{eq:eformalpha2_alpha3},~\ref{eq:eformalpha_alpha1}.

In the high $\widetilde{\mu}_{Se}$ limit (see Fig.~\ref{fig3.ps},
top part), the
 defected interfaces are favored over the mixed ones for almost the whole
 range of variation of $\widetilde{\mu}_{Zn}-\widetilde{\mu}_{Ga}$.
At variance,
 in the  limit of high  $\widetilde{\mu}_{Zn}-\widetilde{\mu}_{Ga}$
 (see Fig.~\ref{fig3.ps},
bottom part)
 all the defected interfaces turn out to be unstable
 over the undefected mixed-ones, irrespectively on the variation of
 $\widetilde{\mu}_{Se}$;
 in the limit of low $\widetilde{\mu}_{Zn}-\widetilde{\mu}_{Ga}$,
 most defected interfaces are favored over the mixed-ones.
Therefore  a low $\widetilde{\mu}_{Zn}-\widetilde{\mu}_{Ga}$
growth condition should favour
   the formation of defected interfaces.

For comparison with experiments, we
observe that ZnSe is usually grown over a GaAs buffer, in a MBE
chamber. Therefore the physically independent
 variables are $\mu_{Zn}$ and $\mu_{Se}$.
As far as $\mu_{Se}$ is concerned, the uppermost value corresponds
to Se-rich condition: in case of evaporation in UHV, like in the
MBE chamber, this would be achieved with a low Zn/Se
beam pressure ratio and consequently a high Se flux, with possible
formation of cluster with a bulk-like crystal
structure from the Se atoms deposited on surface.
The physical interpretation of the other variable,
$\mu_{Zn}-\mu_{Ga}$, is less straightforward. Moreover, in a MBE
chamber, $\mu_{Ga}$ can be considered as a constant as only Zn and
Se atoms come from the effusion cells. Therefore the uppermost value of
$\mu_{Zn}-\mu_{Ga}$ should correspond to \emph{rich} Zn condition
(and viceversa). This interpretation in consistent with the above
discussion about the dependence of formation energy on the basic
variables. We also point out that, in principle, a different
choice of basic variables is possible but this would not
have affected  the main results. Our choice is the
simplest one.

 We observe that 2V, 2V-swap,
1HV, 1HV-swap configurations have the same linear dependence on
$\widetilde{\mu}_{Zn} -\widetilde{\mu}_{Ga}$ and
$\widetilde{\mu}_{Se}$ so that
 we can easily compare their  relative stability in the whole range
 of variability of thermodynamic variables.
 Their relative order, from the most stable,
 is: 1HV, 2V, 2V-swap, 1HV-swap.
 If we define an average number
of ''wrong bonds`` (including the \emph{fictitious} bonds defined
with the vacancies)  per anion in the supercell, we can
empirically observe that (at least for the
  relative stability among these structures) the most stable
 interface is the one with the lowest
  number of Se-Ga and As-V bonds and the highest number of Se-V bonds.
On the other hand, 1V and 1V-swap are stoichiometric but differ for the
presence of
 Zn-As bonds in the latter. As mentioned before, the 1V-swap configuration
 is unstable with respect to 1V, in the whole range of chemical potentials.
This result is consistent with the experimental indication of
Ga--Se bonds but no Zn--As bonds and with the results of
Ref.~\onlinecite{Farrel} which is however limited to undefected
interfaces and does not take into account the charge neutrality
prescription.

Finally we caution the reader that we have just considered trends
in relative stability among different interfaces only on the
basis of thermodynamic arguments, and therefore the high relative stability
of the interface with unrealistic concentration of vacancies
has not an absolute validity. Kinetic effects should
play an important role on stability and in particular could
 hinder the formation of
 interfaces with unrealistic vacancy concentration (see also
discussion in Section ~\ref{VBO}). Taking into account kinetic
effects would go beyond the scope of the present study.
\section{Electronic properties}
\subsection{Density of States}
The presence of vacancies induces in general strong perturbations
on the electronic states due to the large fluctuations of the
crystalline field and may induce localized states at different
energies. As an example, we report in Fig. 4 the total Density Of
States (DOS)
 and the atomic Projected Density Of States (PDOS) on
different atomic layers for the 1HV  and 1HV-swap structures.
 In particular, the PDOS is shown
from the topmost to the bottom panel for the sequence of atomic
planes along the growth direction from the ZnSe side to the GaAs
side through the plane with vacancies. It is also shown the
''excess'' DOS (black area), defined as the difference between the
PDOS and  the corresponding bulk DOS, if positive. The vertical
arrows in each panel denote the energy position of the cation and
anion $s$ states.

From the total DOS (topmost panels) it is evident that the main difference
between the two structures with a high vacancy concentration
in the interface region is
the density  of states in the forbidden energy gap: the structure
1HV is metallic, whereas the 1HV-swap structure maintains the semiconducting
character of its constituents.
Atomic relaxations tend to reduce the effects and to clean the gap,
but not completely, as it can be seen comparing the two topmost panels
for both 1HV and 1HV-swap structures
 (solid/dashed  line  for relaxed/unrelaxed structure)

Looking at the PDOS, it is clear that the gap states in the total
DOS of the  1HV structure originate mainly from states localized
in the interface region where the atomic species have ''wrong''
chemical bonds, with the major contribution deriving  from the Ga
layer sandwiched between  two Se layers  and to the Se atoms
bonded to Ga atoms and vacancies.
An effect of the strong local
fluctuations of the crystalline potential in the interface regions
is the variation in the energy position of the s-Se peaks with
respect to their value in the bulk region and variation in the
shape of the PDOS of Ga atoms bond to Se and As (see the arrows).

By inspection of the PDOS in the bulk sides we can see
a large  difference in the relative positions of the bulk $s$-Se and $s$-As
peaks across the interface in the 1HV and 1HV-swap structures:
in 1HV, the $s$-Se peaks are lower with
respect to the $s$-As peaks by about 4 eV, whereas
 in the 1HV-swap structure this energy difference is strongly reduced and
almost vanishing. This is related to the different ionic interface
dipole due to the atoms swap, sketched in Fig. 1, and results also
in different bands alignments, as discussed in the following
Subsection.

Some of the effects of the fluctuations of the crystalline field
can be explained with a simple model accounting for the ''wrong''
chemical bonds.
We characterize these ``wrong'' bonds by defining the excess average ionic
charge of the nearest-neighbors ($NN$), $\Delta z_{NN}$,\cite{prof}
averaging also over the  vacancies: if $\Delta z_{NN}$ is
positive (negative), the higher (lower) it is, the more
(less) attractive it is the local potential with respect  to
a bulk environment.

In 1HV structure, the As atoms close to the plane of vacancies
have $\Delta z_{NN}$=$-$0.75:  they are in a crystalline potential less
attractive than in the bulk and their valence band states are shifted
upwards in the direction of the
forbidden energy gap. The other atoms in  the interface region
have a positive $\Delta z_{NN}$: $\Delta z_{NN}$ = 0.5, 1.0, 0.25,
0.5, respectively, for the Se layer between Zn and Ga, for
Ga between Se layers, for Se between Ga layers, and
for  the Ga
plane with vacancies. A crystalline potential more attractive than
the bulk one pushes down in energy the states from the conduction
band. This roughly explains the origin of
the gap states in  this structure.

In the 1HV-swap structure, the local fluctuations of the crystalline field
are weaker.  Se atoms between the Zn  and the  Ga$_{0.5}$ layer with vacancies
show electronic states not very different with respect to the bulk;
and Zn atoms nearest to the interface region (with $\Delta z_{NN}$=0)
also show features similar to the bulk. The Se atoms sandwiched between two
Ga planes  (with $\Delta z_{NN}$=0.25), strongly differ from the bulk case but
not in the region of the energy gap.

Extending our analysis to the other defected interfaces studied here,
we notice that the presence of electronic states within the energy gap
is a common feature, and the  1HV-swap structure is an exception.

\subsection{Band alignments}\label{VBO}

We summarize schematically  in Fig.~\ref{fig5.eps} the band
alignments calculated for the selected structures
 considered here. We follow the approach of
 Ref.~\onlinecite{review}, splitting the band offset into two
contributions:
 $VBO=\Delta E_{v}+\Delta V$, where $\Delta E_v$ is the band structure term,
i.e. the energy difference between the relevant valence band top edges
of the two materials
measured with respect to the
 average electrostatic potential in the corresponding bulk crystal and
$\Delta V$ is  the electrostatic potential lineup  containing all
interface-specific effects and extracted from supercell
calculations. The potential lineup in turn includes a contribution
related to the ionic dipole and its screening (indicated as
$\Delta V_{hetero}$ in Ref.~\onlinecite{review}) plus a term which
is purely electronic and independent on interface details.   The
spin orbit effects are added  \emph{a posteriori} using
experimental data.

As predicted for the  heterovalent heterostructures,
 the band alignments are strongly dependent on the
specific interface morphology.
This is particularly true
in the occurrence of vacancies, as it can be easily seen
by inspection of  Fig. 5:
for undefected (without vacancies) interfaces
 the maximum variation is of the order of  1 eV
(compare 1C and 1A cases),\cite{unstable4,kumar,FunatoJAP1999}
whereas it is  much larger in case of vacancies by changing their
concentration and position, the extreme variation (of about 4 eV)
being found for
 1HV and 1HV-swap structures, where the alignment
changes from the so-called ``broken gap'' type (1HV) to ``staggered''
(1HV-swap).
We point out that such a huge variation is associated to unrealistically
high local concentration of vacancies. In real samples we expect
 vacancies  diluted  over a larger interface region.

In the spirit of
{\em linear response theory} (LRT)~\cite{reconstr1,review,Franciosireport}
the variation of band offset in heterovalent heterojunctions
 has been rationalized and quantitatively explained in terms of  the
line-ups  associated to the different interface ionic charge
distribution and the corresponding  electronic screening, i.e. of
$\Delta V_{hetero}$. Neglecting details such as optimized atomic
positions, simple calculations based on elementary electrostatics
and LRT give $\Delta$V$_{hetero}$=${\pi e^2/ 2 a_0
\langle{\epsilon}\rangle} \approx 0.5$ eV for 1C, 1V and 2V, the
opposite for 1A, $\Delta$V$_{hetero}$=0 for 2CA-ZnSe and 2CA-GaSe,
and $\Delta$V$_{hetero}$=${9\pi e^2}/{2
a_0\langle{\epsilon}\rangle}\approx$ 4 eV for 1HV and
$\Delta$V$_{hetero}$=$-{3\pi e^2}/{2
a_0\langle{\epsilon}\rangle}\approx$ $-$0.33 eV for 1HV-swap,
where $a_0$ is  the lattice constant and
$\langle{\epsilon}\rangle$  a proper average~\cite{epsilonaverage}
between the  dielectric constant of GaAs and ZnSe (which are 10.86
and 5.7 for GaAs and ZnSe respectively~\cite{handbook}). By
comparing the LRT predictions with the full self-consistent
calculations reported in Fig. 5 we can comment about the validity
of the LRT: its predictions are quantitatively correct
for undefected structures, but
fails with discrepancies up to 1 eV in case of vacancies (LRT
predicts band offset 1V and 2V equal to 1C, and 1V-swap equal to
2V-swap). It is not surprising: the LRT better holds in case of
interface configurations without vacancies since the perturbation
due to the presence of the interface is related to rather small
differences   between atoms (e.g. Zn vs. Ga, Se vs. As), and it is
expected to be less accurate in case of strong perturbations such
as vacancies.

The  measured variation of the band offset in real
samples obtained with different
growth procedures is much lower than what found here
for different vacancy configurations:
it is limited to about 0.6 eV, as indicated
by the range of the experimental data
in the right hand side of Fig. 5 with the label ``Exp(2)''.
We show separately  the right hand side of Fig. 5 with the label
``Exp(1)'' the experimental values from
Ref.~\onlinecite{ColliJAP2004}, ranging from 0.60$\pm$0.05 eV to
0.72$\pm$0.10 eV for the two samples with lowest SF density, whose
composition of the interface region is indicated to be 45 at\% Se,
34 at\% Zn, 21 at\% Ga and 50 at\% Se,  40 at\% Zn, 10 at\% Ga
respectively.
A common indication extracted from the experiments is a type-I or
``straddling'' band alignment (the band edges in the GaAs slab are
both {\em within} the band gap of ZnSe): remarkably, this is
qualitatively compatible with most configurations studied here,
including some with vacancies.

It is not possible to extract any precise indication on
the morphology of real samples from the comparison between
experimental results and numerical predictions: there are
many different configurations which can give the same band offsets
and, moreover, relative stability crucially depends of the
kinetics of the growth process. However, we can conclude that the
occurrence of vacancies is compatible with the experimentally
measured band offsets, and  the most likely configurations in real samples
seems to be
characterized not only by charge neutrality but also by vanishing
ionic dipoles.

\section{Conclusion}
We have studied with accurate ab-initio calculations
several selected interface configurations for
the  ZnSe/GaAs(001) heterojunctions satisfying the charge neutrality condition,
 including also defected configurations with cation vacancies.
We have shown that in some particular thermodynamic conditions
(basically \emph{rich}-Se chemical potential) the formation of
defected interfaces with cation vacancies is even favoured over
the simpler, undefected, unstrained,
 anion- and/or cation-mixed  interfaces. We would like to note
 that one of the procedure leading to  interface with low SF's density (and
 vacancy at interface) involve fabrication of ZnSe buffer layer at
 interface in highly Se-rich condition.\cite{ColliJAP2004}
The band alignments are strongly dependent on the particular
interface morphology, in particular in the occurrence of
vacancies; on the other hand, different morphologies can
correspond to the same or to similar band offsets, so that a
unique correspondence between morphology, band offset, relative
thermodynamic stability
 cannot be established.
Nevertheless, the predicted stability of some defected interfaces
and the compatibility of predicted band alignments with
measurements support the experimental evidence of (Zn,Ga)Se
defected compounds in high-quality ZnSe/GaAs(001) heterojunctions
with low native stacking fault density. The range of measured band
offset suggests that different atoms at interfaces rearrange, with
possible presence of vacancies, in such a way that not only local
charges but also ionic dipoles are vanishing.

%%%%%%%%%%%%%%%%%%%%%%%%%%%%%%%%%%%%%%%%%%%%%%%%%%%%%%%%%%%%%%%%%%%%%%%%%%%%%%
\section{Acknowledgments}

We gratefully acknowledge E. Carlino, A. Franciosi and A. Colli for
fruitful discussions.
Computational resources have been obtained
partly  within the ``Iniziativa Trasversale di Calcolo
Parallelo'' of the Italian {\em
Istituto Nazionale per la Fisica della Materia}
(INFM) and partly within the agreement between
the University of Trieste and the
Consorzio Interuniversitario CINECA (Italy).

\newpage

\begin{figure}[!hbp]

\caption{Composition and bare ionic charge profiles (in unit of
electronic charge per zincblende unit cell) in the interface
region of the defected heterostructures 1HV, 1HV-swap, 1V,
1V-swap, 2V, 2V-swap along the (001) growth direction. The ionic
charge is averaged over pairs of adjacent (001) atomic layers and
normalized to the bulk zincblende unit cell. Interface cationic
planes are emphasized in boldface.} \label{fig1.ps}
\end{figure}
%%%%%%%%%%%%%%%%%%%%%%%%%%%%%%%%%%%%%%%%%%%

\begin{figure}[!hbp]
\caption{Interplanar distances (in \AA) for the different structures
considered including cation vacancies.
The average value of the
interplanar distances are indicated by closed symbols,
and the error bar indicate the range of local variation due to the
possible buckling of the atomic planes.
}
\label{fig2.ps}
\end{figure}
%%%%%%%%%%%%%%%%%%%%%%%%%%%%%%%%%%%%%%%%%%%%%%%%%%%%%%%%%%%%%%%%%%%%%%%

%%%%%%%%%%%%%%%%%%%%%%%%%%%%%%%%%%%%%%%%%%%
\begin{figure}[!hbp]
\caption{ {Top}: Formation energy [in eV/1$\times$1 surface] of
the mixed and defected interfaces as a function of the difference of
the Zn and Ga chemical potentials,
 for lower(upper) limit of $\widetilde{\mu}^{Se}$.
Dotted lines refer to the undefected 1C, 1A, 2CA-ZnAs, 2CA-GaSe
interfaces, which appear
degenerate on this energy scale.
Hatched areas indicate the range of interface formation energy
for interfaces with the same stoichiometry:
right (left) hatched areas represent the range of variation of
formation energy for 1HV, 1HV-swap, 2V, 2V-swap (1V, 1V-swap).
Those with the lowest
formation energy are 1HV-swap and 1V for the two families
respectively. {Bottom}: as
before, as a function of the difference of the Se chemical
potential for the lower (upper)  limit of
$\widetilde{\mu}^{Zn}-\widetilde{\mu}^{Ga}$. The chemical potentials
have been rescaled with respect to their correspondent bulk value
(see text).} \label{fig3.ps}
\end{figure}

%%%%%%%%%%%%%%%%%%%%%%%%%%%%%%%%%%%%%%%%%%%%%%%%%%%%%%%%%%%%%%%%%%%%%%%

\begin{figure}[!hbp]
\caption{Total DOS (upper panels) and atomic Projected Density of
States over layers along the growth direction for  the 1HV (panels
a) and 1HV-swap (panels  b)  interface. The ''excess`` DOS is
shown as black area (see text). Arrows indicate the energy
position of the cation and anion $s$ states in each layer.
Energies are referred to the Fermi energy for 1HV and to the
Valence Band Maximum for 1HV-swap. Solid/dashed  line in the
topmost panels
 are for relaxed/unrelaxed structure.} \label{fig4.ps}
\end{figure}

\begin{figure}[!hbp]

\caption{Scheme of the band alignments  for the different
interfaces investigated. The relative energy scale is referred to
the band edges of GaAs, which are indicated on the left, and the
relative position of ZnSe is then indicated for the different
morphologies. The horizontal dotted lines refer to GaAs band
edges. The label ``Exp(1)'' refers to experimental values in
Ref.\onlinecite{ColliJAP2004} and ``Exp(2)'' to all the other
measurements reported in the literature\cite{unstable4}.}
\label{fig5.eps}
\end{figure}

\newpage
%%%%%%%%%%%%%%%%%%%%%%%%%%%%%%%%%%%%%%%%%%%

\clearpage
\includegraphics[scale=.7,angle=0]{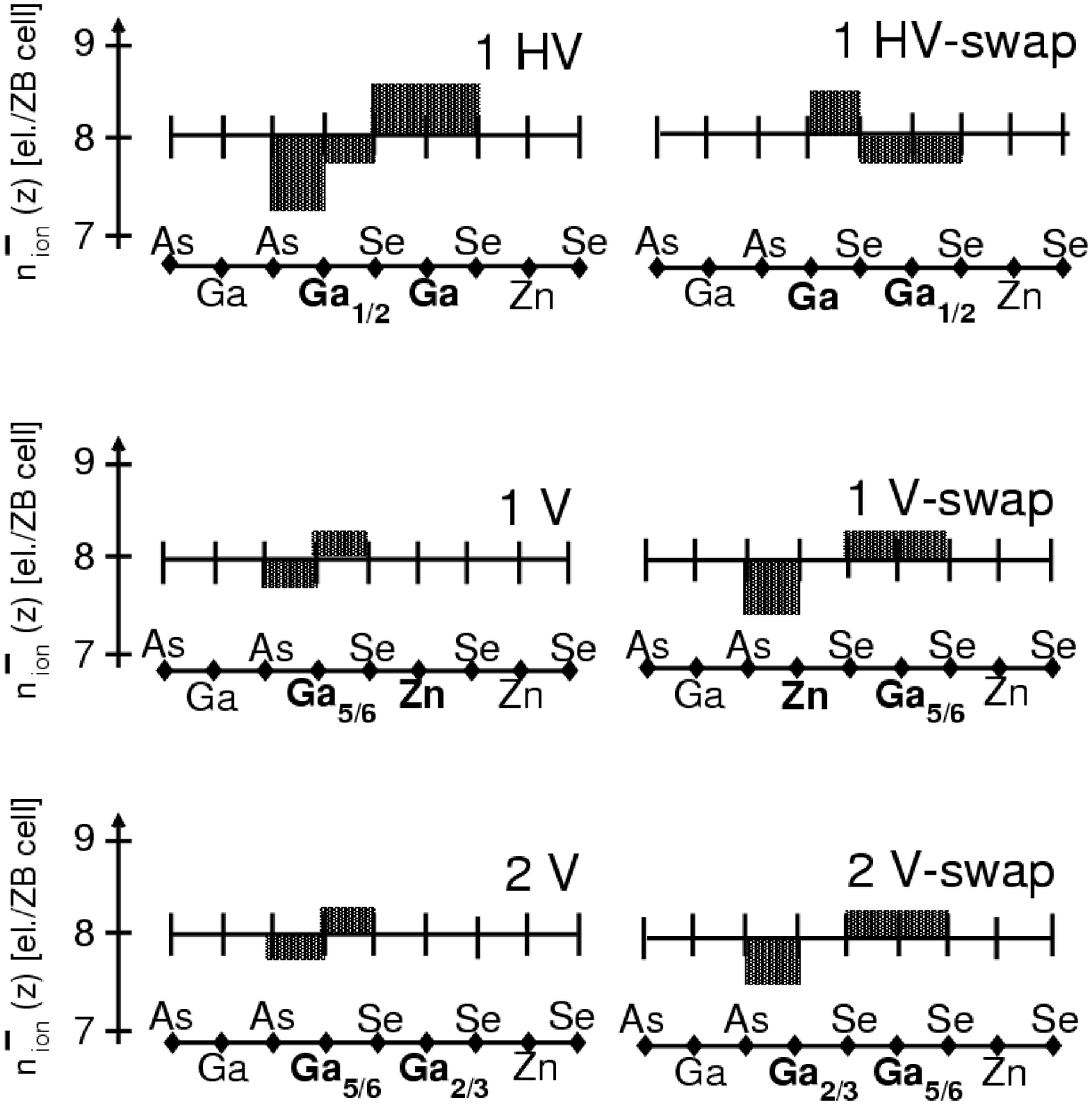}
\\Fig. 1
\clearpage
\includegraphics[scale=0.8,angle=0]{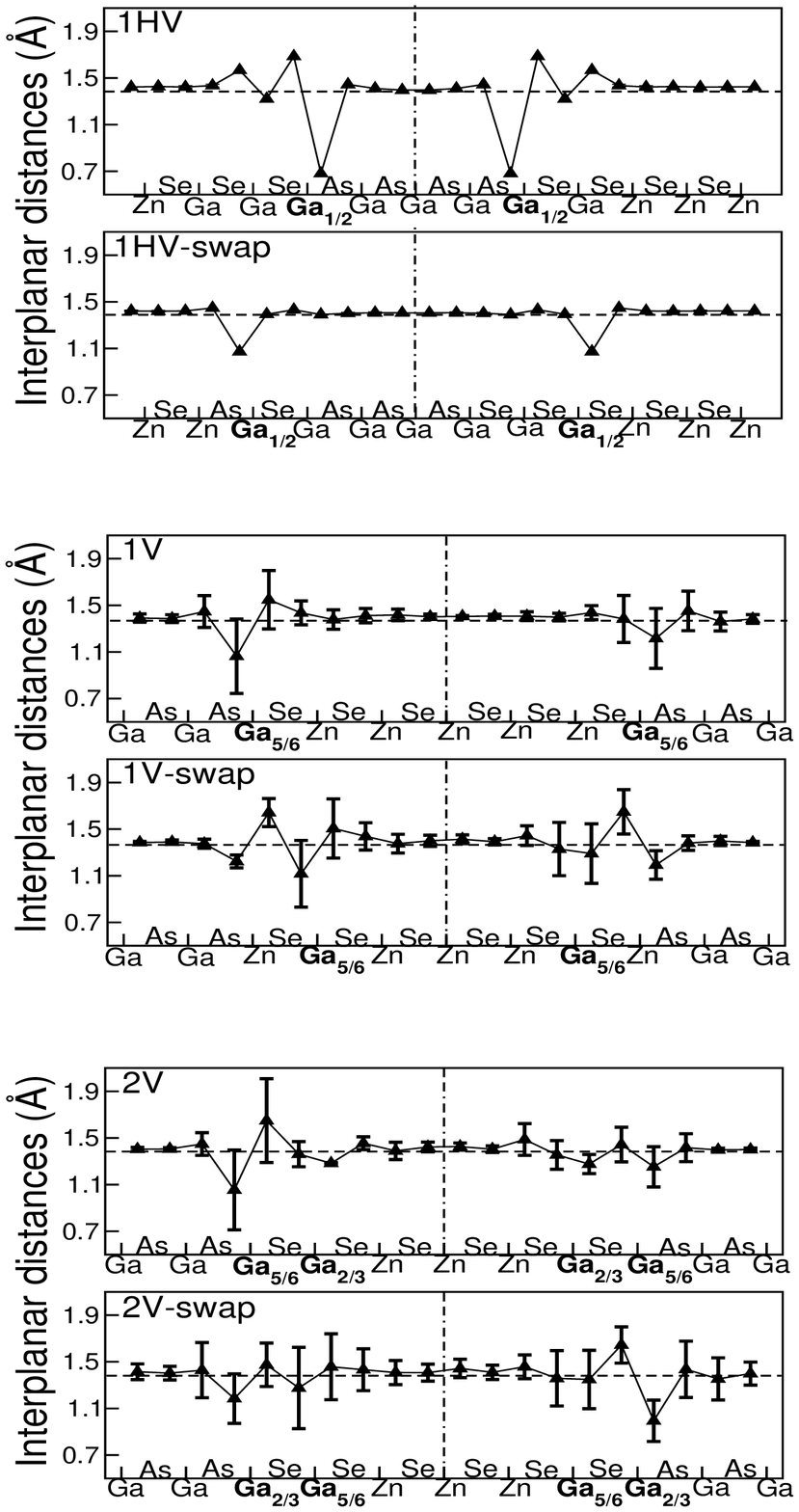}
\\Fig. 2
\clearpage
\includegraphics[scale=0.8,angle=0]{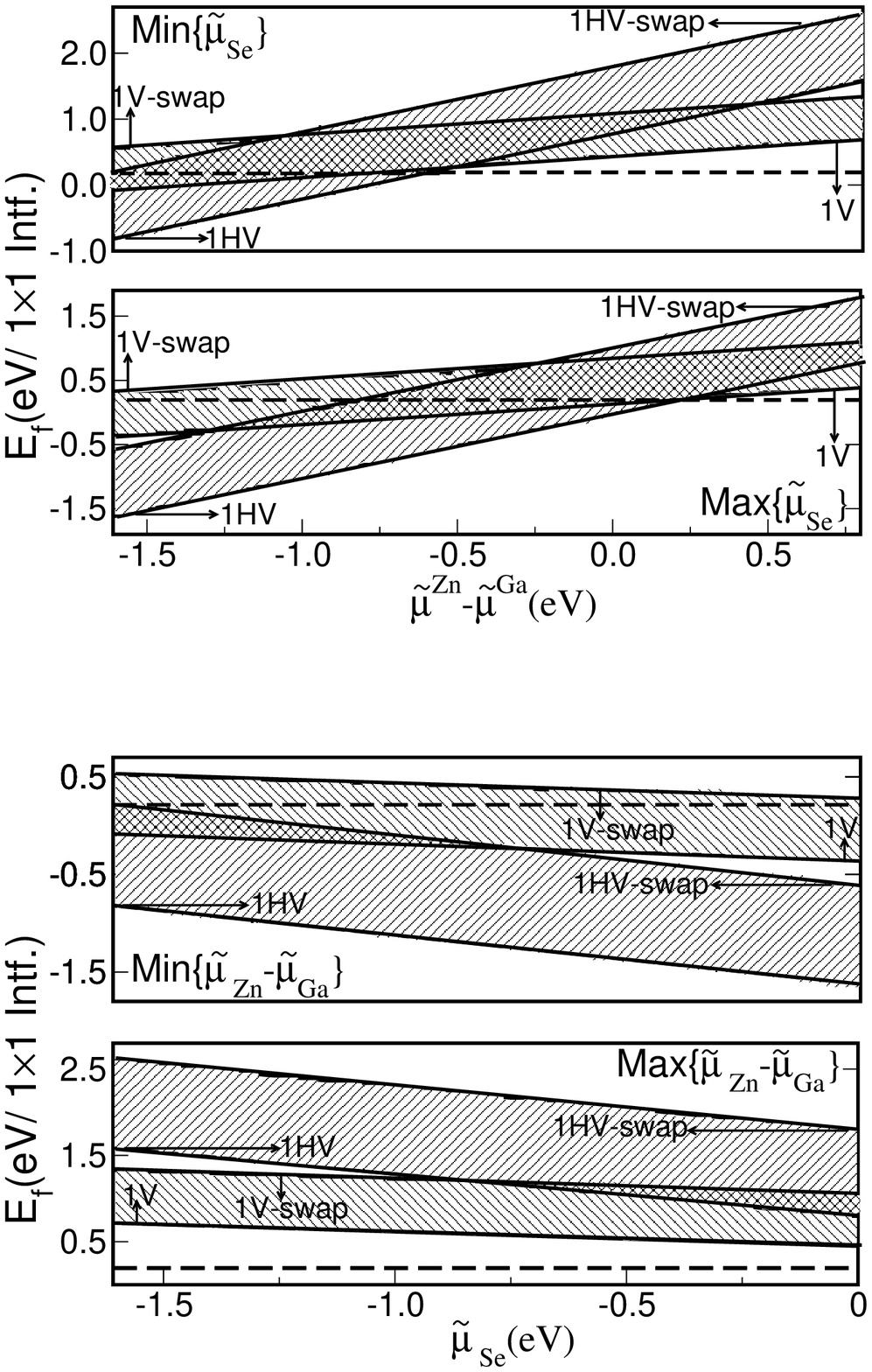}
\\Fig. 3
\clearpage
\includegraphics[scale=0.65,angle=-90]{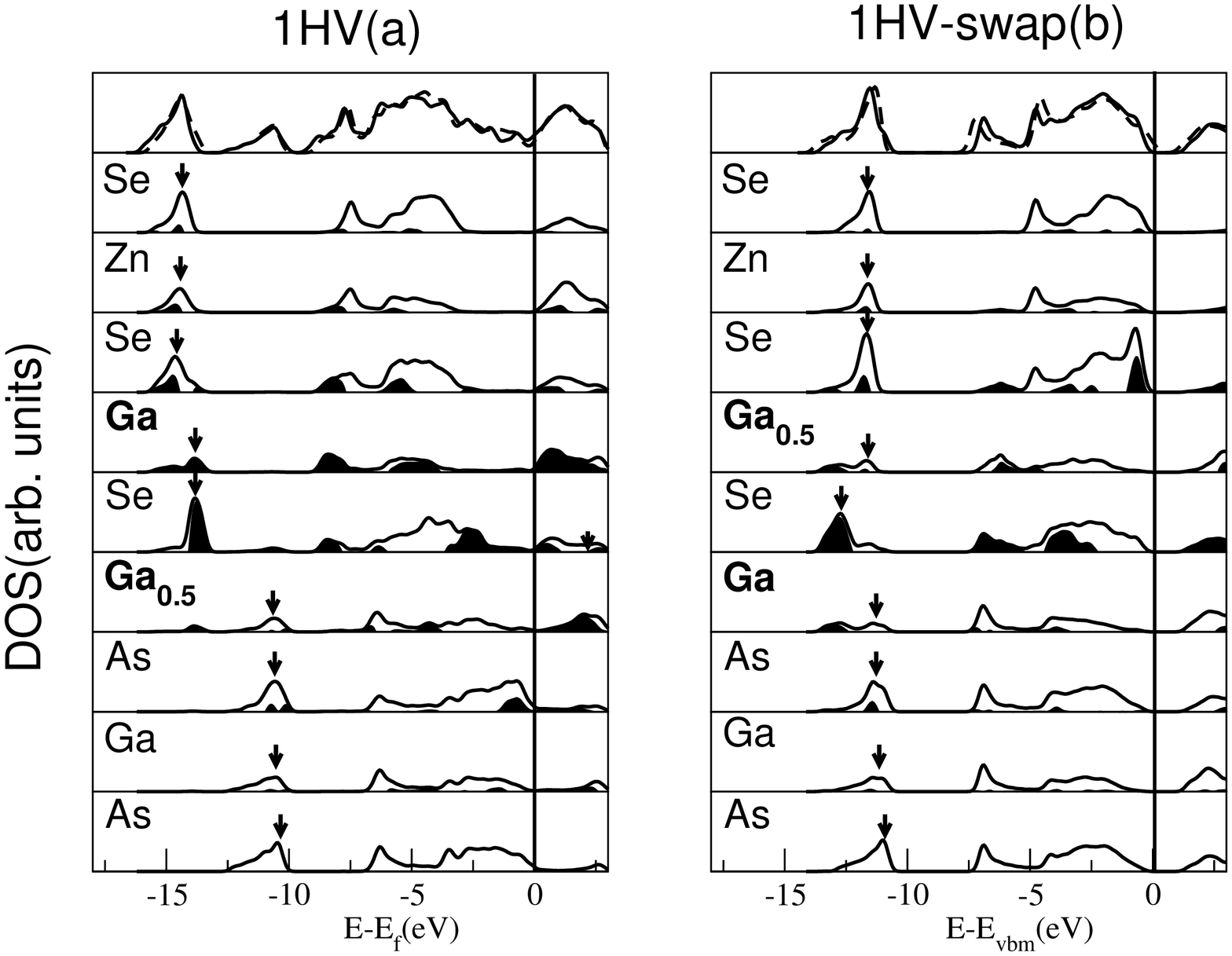}
\\Fig. 4
\clearpage
\includegraphics[scale=0.65,angle=-90]{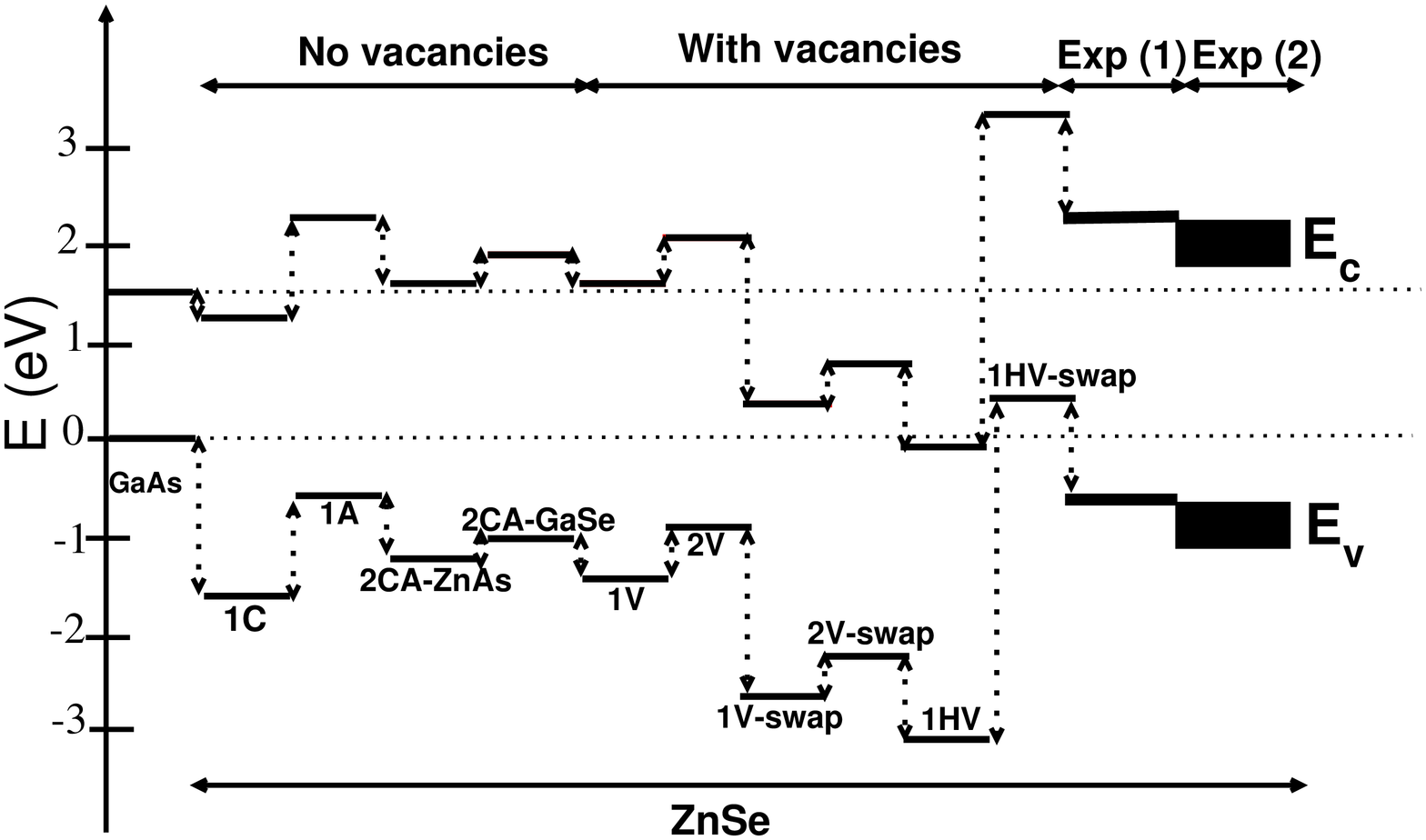}
\\Fig. 5

\begin{thebibliography}{}
\bibitem{forces}
For the optimization of atomic position we require
Hellmann-Feynman forces  smaller then 0.02 eV \AA$^{-1}$.
\bibitem {note1} The condition must be strictly expressed
in terms of ionic charges, since the electronic screening is not
completely achieved within the interface region.
\end{thebibliography}
\end{document}